\pdfoutput=1

\documentclass[11pt]{article}

\usepackage[final]{coling}

\usepackage{times}
\usepackage{latexsym}
\usepackage{float}
\usepackage{amsmath}
\usepackage{amsthm}
\usepackage{csquotes}
\usepackage{caption}

\usepackage{amsfonts}
\usepackage{amssymb}
\usepackage{svg}
\usepackage{listings}
\usepackage{xcolor} 
\usepackage{soul}

\usepackage{enumitem}
\setlist{nolistsep}

\theoremstyle{plain}
\newtheorem{assumption}{Assumption}

\newtheorem{claim}{Claim}

\usepackage[T1]{fontenc}

\usepackage[utf8]{inputenc}

\usepackage{microtype}

\usepackage{inconsolata}

\usepackage{graphicx}

\title{MST-R: Multi-Stage Tuning for Retrieval Systems and Metric Evaluation}

\author{Yash Malviya$^*$, Karan Dhingra$^*$ \and Maneesh Singh \\
         Indic aiDias Team (Independent Researchers) \\
         \texttt{\{yashmalviya98,kdhingra307\}@gmail.com} \\
         \small{\textbf{Correspondence:} \href{mailto:dr.maneesh.singh@ieee.org}{dr.maneesh.singh@ieee.org}}
}

\begin{document}
\maketitle

\def\thefootnote{*}\footnotetext{These authors contributed equally to this work.}\def\thefootnote{\arabic{footnote}}

\begin{abstract}
Regulatory documents are rich in nuanced terminology and specialized semantics. FRAG systems: Frozen retrieval-augmented generators utilizing pre-trained (or, frozen) components face consequent challenges with both retriever and answering performance. We present a system that adapts the retriever performance to the target domain using a multi-stage tuning (MST) strategy. Our retrieval approach, called MST-R (a) first fine-tunes encoders used in vector stores using hard negative mining, (b) then uses a hybrid retriever, combining sparse and dense retrievers using reciprocal rank fusion, and then (c) adapts the cross-attention encoder by fine-tuning only the top-k retrieved results. We benchmark the system performance on the dataset released for the RIRAG challenge (as part of the RegNLP workshop at COLING 2025). We achieve significant performance gains obtaining a top rank on the RegNLP challenge leaderboard. We also show that a trivial answering approach {\em games} the RePASs metric outscoring all baselines and a pre-trained Llama model. Analyzing this anomaly, we present important takeaways for future research. We also release our \href{https://github.com/Indic-aiDias/MST-R}{code base}\footnote{\url{https://github.com/Indic-aiDias/MST-R}}.
\end{abstract}

\section{Introduction}
Automated Q\&A systems hold tremendous potential in not only improving access to, and comprehension of regulatory obligations, but also help organizations achieve regulatory compliance with reduced costs and latency. Currently, compliance workflows are largely manual and organizations need to employ a large number of costly subject matter experts. High recall is especially critical in this domain, as the cost of false negatives i.e. missing crucial regulatory information can lead to severe financial penalties, legal repercussions, and reputational harm. Retrieval-augmented generation (RAG) offers a promising solution but their performance falls short when FRAGs (frozen RAGs or RAGs with pre-trained (frozen) components) are directly applied since regulatory documents utilize specialized, domain-specific terminology and nuanced legal semantics. Domain-specific adaptations are needed to make these systems viable. 

This paper primarily focuses on the retriever part of the system, presenting a simple domain adaptation approach to significantly improve the performance of the retriever part of the system by fine-tuning on the target domain. Our contributions include: (a) MST-R A multi-stage retrieval system domain adapted using a multi-stage fine-tuning approach. (b) State of the art retrieval performance on the RIRAG challenge with a improvement of 12.1\% in $Recall@10$ and 23\% in $MAP@10$ compared to the BGE baseline from \citealt{gokhan2024regnlp}. (c) Analysis of the RePASs metric with a solution that {\em games}, with important takeaways.

\section{Prior Work}
Passage retrieval is a critical step in RAG systems. Early methods relied on sparse representations such as TF-IDF and BM25 \cite{bm25}. Dense encoders\cite{bge, e5} using late interaction\cite{colbert}, enable document embedding caching. In contrast, approaches such as \cite{msmarco, qa_system_1, hyrr} use query-document interaction for nuanced semantic alignment but incur higher computational costs due to per-sample processing. Recently, hybrid search algorithms \cite{bge_bm25, rrf} combining lexical patterns with semantic relationships between queries and passages have emerged. Advanced retrieval systems like Re2G \cite{glass2022re2g} employ multi-level architectures to optimize performance and efficiency. Inspired by the above, we propose a multi-level architecture which fuses the results of a variety of approaches from the literature to leverage their complementary strengths. 

The retrieval stage in QA systems is typically evaluated using metrics like Recall@k and Mean Average Precision (MAP@k). For answer quality, metrics such as BLEU and ROUGE focus on n-gram statistics but neglect semantic equivalence in abstractive generation. Semantic-focused metrics \cite{oags, ragas, yue2023automaticevaluationattributionlarge, Laban2022SummaCRN} emphasize alignment and coherence but lack comprehensive coverage, especially in regulation domains. To address this, RePaSs \cite{gokhan2024regnlp} enforces coverage of all obligations within relevant passages, ensuring both relevance and regulatory compliance.

\section{Methodology}
\label{sec:approach}
\begin{figure}[!htb]
    \centering
    \includesvg[width=1.0\linewidth]{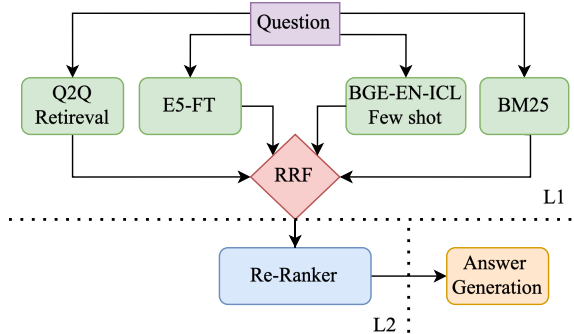}
    \caption{Multi Stage Tuning System Architecture}
    \label{fig:system-design}
\end{figure}

Our system, depicted in Fig. \ref{fig:system-design}, is a simple, single pass, feed-forward RAG system, similar to \citealt{glass2022re2g}. A two-stage retriever selects the relevant passage(s) to provide to a Q\&A module. For the latter, we use a pre-trained\footnote{Arguably, domain adaptation of the Q\&A module will improve the performance of our system further.} prompt-tuned LLM.

\subsection{Passage Retrieval}
The retriever part of our system comprises of two stages (levels). Level 1 (L1) uses the user query to return a ranked list of results. We use a custom hybrid index adapted to the regulatory domain. Level 2 (L2) employs a reranker using a costly cross-attention mechanism to perform a fine-grained analysis of the relevance of the retrieved answer to the query. We adapt the reranker to the target domain by fine-tuning it on the results retrieved from Level 1. This two-level approach allows us to obtain a good trade-off between cost and performance. 

\subsubsection{Level 1: A Domain-Adapted Hybrid Retriever}
Dense retrievers utilize ANN search in semantic spaces where document chunks are embedded using a pre-trained DNN encoder. Dense retrievers have shown excellent performance invariant to the exact formulation of the text as long as the meaning is retained. Fine tuning such retrievers (not training from scratch) is expected to result in partial domain adaptation. To cover for such a shortfall, we also utilize sparse retrievers which use lexical similarity between the question and document chunks. Hence, we design Level 1 to comprise the following retrievers. 

(a) \textbf{BM25}: is a sparse retriever model leveraging a modified TF-IDF formulation to retrieve passages, and capture lexical similarity between questions and passages. We expect its performance to be covariant with the domain. 
(b) \textbf{BGE-EN-ICL}: is a dense retriever model allowing for prompt-based few-shot learning. We adapt it to the target domain by conditioning it on five random samples from the ObliQA dataset.
(c) \textbf{E5-FT}: is a version of the E5 dense retriever model\cite{e5} created by fine-tuning on the ObliQA dataset, thus adapting it to the target domain. We used contrastive learning with the triplet loss\cite{schroff2015facenet} and performed online hard mining\cite{shrivastavaCVPR16ohem}. We provide more details in the Appendix \ref{appendix:l1_finetuning}.
(d) \textbf{Q2Q}\cite{q2q}: A retriever is a map between the space of queries and the space of relevant passages. A new query is an index in this map. Assuming standard regularity conditions, an ANN can be used to retrieve similar queries previously encountered by the system (e.g. questions in the training data). Ground truth passages corresponding to the most relevant previously seen questions are then used. For encoding, MPNet\cite{mpnet} model was fine-tuned similar to E5-FT.

To combine the results from the above four retrievers, we used reciprocal rank fusion (RRF) \citep{rrf}. The RRF score $S(k)$ of the $k^{th}$ passage is given in terms of its ranks $R(k, i)$ by the $i^{th}$ ranker, by,
$    S(k) = \sum_{i=1}^{n} \frac{1}{R(k, i) + \beta} $. Please refer to the Appendix \ref{appendix:rrf} for an analysis of this design choice.

\subsubsection{Level 2: Domain-Adapted Reranker}
For re-ranking, the \emph{ms-marco-MiniLM-L-6-v2} model\cite{msmarco, minilm}, a cross-encoder trained on the MSMARCO dataset \citep{bajaj2016ms} for document re-ranking, is fine-tuned by replacing its task head with a binary classification head for relevance prediction\footnote{The choice of task head was motivated by experimental exploration.}. The probability of belonging to the {\em relevant} class was utilized as the reranking score.

The fine-tuning dataset is constructed using relevant passages from the training ground truth combined with negative sampling. Hard negative samples are selected from the top-$K$ documents of different $L1$ retrievers that are not relevant in the ground truth. Easy negative samples are generated through random sampling passages from the corpus.

\subsection{Answer Generation}

Ideally, the answering LLM should also be adapted to the target domain. In this work, we have just used Llama3.1 Instruct 8B\cite{llama3.1} with the prompt mentioned in \cite{gokhan2024regnlp} \& Appendix \ref{appendix:prompts}. Since both (a) better models can be used, and (b) they can be adapted to the target domain, the performance obtained here should be considered as minimum achievable performance. In addition, we evaluated two other {\em default} strategies of passing the input directly to the output without using an LLM to formulate an answer\footnote{Refer to Section \ref{sec:analysis}.}. Thus, in total, we evaluated three strategies - (a) \textbf{LLM - RegNLP Prompt}\cite{gokhan2024regnlp} using Llama3.1 Instruct 8B\cite{llama3.1}: Appendix \ref{appendix:prompts}. (b) \textbf{Passage Concatenation} (PC): simply concatenates the retrieved passages and provisions them as the answer. (c) \textbf{Single line} (SL): removes the sentence terminators from the above, converting the entire answer into a single line answer.

\section{Evaluation}
We now present the results of the proposed system on the ObliQA dataset \cite{gokhan2024regnlp}. A simple pre-processing step was applied to remove section headers, etc. by filtering out passages with fewer than 10 tokens.

Following the guidelines in the challenge, we used $Recall@10$ and $MAP@10$ to evaluate retrieval performance and the RePASs metric to evaluate the goodness of the generated answer. 

RePASs metric is defined using entailment, contradiction, and obligatory coverage scores as follows: $RePASs = \frac{ \text{E}_{s} - \text{C}_{s} + \text{OC}_{s} + 1}{3}$. Entailment score, $E_s$ (or Contradiction score, $C_s$) measures whether an answer sentence is entailed (or contradicted) by a retrieved context sentence. $OC_s$ (Obligation Coverage score) measures the percentage of obligations present in the retrieved context that are covered by the answer.

\begin{table}[!ht]
    \centering
    \begin{tabular}{|l|l|l|}
    \hline
        Algorithm & $Recall@10$ & $MAP@10$ \\ \hline
        MST-R (L1+L2) & \textbf{0.8746} & \textbf{0.7601} \\ \hline\hline
        RRF(L1) & 0.832 & 0.6914 \\ \hline \hline
        BGE (5 shot) & 0.7796 & 0.6178 \\ \hline
        BM25 & 0.7611 & 0.6236 \\ \hline 
    \end{tabular}
    \caption{Retrieval performance on the ObliQA dataset. Detailed ablation of L1 is presented in Appendix: \ref{appendix:ablation}.}
    \label{table:results_l1}
\end{table}

{\bf Retrieval results} are presented in Table \ref{table:results_l1}. While the two baselines, BM25 and BGE (with 5 shot) have $Recall@10$ scores of $0.76$ and $0.78$ respectively, L1 level of our system gives a score of $0.83$ - a boost of $6.7\%$ relative to BGE. Incorporating the reranker boosts it to $0.87$, an additional $5.12\%$ relative to L1. 

\begin{table}[!ht]
    \centering
    \begin{tabular}{|p{2.7cm}|l|l|l|l|}
    \hline
        Method & (RePASs,$E_s$,$C_s$,$OC_s)$ \\ \hline
        Llama3.1-Instruct-8B & (0.41, 0.215, 0.091, 0.129) \\ \hline
        Single Line & (0.801, 0.715, 0.098, 0.786) \\ \hline
        Passage Concat & (\textbf{0.947}, \textbf{0.986}, \textbf{0.076}, \textbf{0.932}) \\ \hline
    \end{tabular}
    \caption{Performance of various answer generation strategies on the ObliQA dataset.}
    \label{table:ans_gen_exp}
\end{table}

{\bf Answer generation} performance is presented in Table \ref{table:ans_gen_exp}. Note that the Llama model gives a fairly low score of $0.41$ primarily due to low entailment and low coverage. A better LLM, finetuned on the domain, can arguably improve the performance significantly. This is not yet a part of our study. On the other hand, we noticed something peculiar: the RePASs metric is not a {\em complete} metric. To demonstrate this, we tried two simple baselines: Passage Concatenation ({\em PC}) \& Single Line.  

Astonishingly, {\em PC} achieves a rather high score of $0.947$, $130\%$ relative improvement in the RePASs metric over the Llama model. Even when we convert the entire answer into a single line, we get a rather high performance of $0.801$. These results highlight the limitations of the RePASs metric for in evaluating answer quality.

\section{Analysis: The RePASs Metric}
\label{sec:analysis}

RIRAG uses standard metrics to evaluate the retrieval performance but defines a novel metric called RePASs to evaluate the answer. RePASs is reference-free, using neither the ground truth answer nor the input question. In this section, we analyze the properties of the metric itself.

\subsection{Trivial Optimizers?}
\label{experiment:repass_trivial_solution}
The RePASs metric is optimized by an answer with a high entailment score and obligatory coverage and with no contradiction. Arguably, an answer constructed from concatenating passages meets the above criterion as long as its content is not self-contradictory. Even a meaningless concatenation of all retrieved passages into a single line outscores the Llama model (results in Table~\ref{table:ans_gen_exp}). More detailed analysis is provided in Appendix \ref{optimal-repass}. 

\subsection{Reasoning Context: RePASs-N?}
The RePaSs metric averages over sentence level entailment and contradiction. Stripping source and passage sentences from the surrounding context makes reasoning hard leading to an erratic metric. We investigate below the behavior of the entailment and contradiction scores from NLI models as a function of larger window sizes.

Since the ObliQA dataset doesn't contain GT answers, we used the CNN/Daily Mail dataset \cite{cnn_daily}. We provision the source and GT summaries to the Deberta v3 NLI model\cite{deberta} using the same exhaustive, sliding-window strategy as RePASs but with varying context size N (N+1 sentences). We call this extension to the RePASs metric, RePASs-N. 

\begin{figure}[!htb]
    \centering
    \includesvg[width=0.9\linewidth]{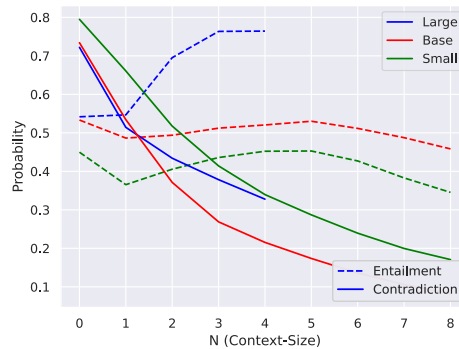}
    \caption{$E_s$ and $C_s$ for RePASs-N. NLI Deberta v3\cite{deberta} used with context length N.}
    \label{fig:ngram_experiment}
\end{figure}

Results are shown in Figure \ref{fig:ngram_experiment}. A reasonable expectation is for the entailment (contradiction) score to increase (decrease) with larger N. Only the \lq Large\rq model is able to adequately reason using the larger context. While more exhaustive experimentation is needed, results with $N=3$ in (Table \ref{table:n-chunk-repass-score}) in the Appendix shows a relative improvement of ~20\% in RePASs-3 metric over RePASs (or RePASs-0), with corresponding relative improvements of ~68\% in entailment and contradiction scores.    

\subsubsection{Better Metrics Needed?}
The existence of trivial optimizers for RePASs-N metrics posits the need for additional metrics which ensure relevancy, accuracy and succinctness of the answer and whether it conforms to good form (style, morphology, etc.). The reader is encouraged to take a look at the good body of research on metrics\cite{ragas}. An emergent trend is to use \lq LLM-as-a-Judge\rq metric to evaluate answer quality. \cite{zheng2023judging} proposes a reference-free metric requiring the original question and the generated answer. 

We evaluated {\em Passage Concat} and {\em Llama} answers on this metric. Specifically, we used the prompt in \cite{roucherllmjudge} to ask the LLM to judge whether the answer is relevant, direct, detailed, and addresses all the concerns in the question. Prompt details are provided in \ref{appendix:prompts}. Results show that {\em Passage Concat} has a high RePASs score (0.947) but a lower LLM-as-a-Judge score (2.823). Conversely, answers from Llama have a low RePASs score (0.41), but a high LLM-as-a-Judge score (3.91). The above exploration shows that better metrics are needed to evaluate the goodness of the answer in the regulatory domain: either a single more comprehensive metric, or a list of metrics covering various aspects of answer goodness.

\section{Conclusion}
The regulatory domain presents significant challenges due to its complex language and contextual requirements needing good domain adaptation strategies. We presented a domain-adapted, multilayered retrieval system showing significant performance gains. While we didn't adapt the performance of the answer generator (leaving it for future work), we present the need for better evaluation metrics as a precursor to engineering better answering models.

{\bf Limitations}: Considerable research is needed to engineer per-formant answering systems for regulatory domains. While we present an engineered system, it has not been comprehensively evaluated to be deployment-ready. On the other hand, our paper is indicative of the research needed for creating such systems.

\bibliography{coling_latex}

\appendix
\section{Appendix}

\subsection{Reciprocal Rank Fusion}
\label{appendix:rrf}

To understand this design choice, note that (a) RRF is dependent on only the ranks of the retrieved results, allowing us to avoid calibrating and fusing distances from different spaces, and (b) RRF can be thought of as a naive fusion model where document relevance, $rel$, decays exponentially with the reciprocal of its retrieval rank $\pi$: $P(rel|\pi) = exp(1/(\pi+\beta))$. Results from $K$ retrievers
can be fused using a naive notion of conditional independence:  $ logP(rel|\{\pi_i\}_{i=1..K}) \approx log \prod_{i=1}^K p(rel|\pi_i) = \sum_{i=1}^K \frac{1}{\pi_i + \beta}$. In our implementation, we use four retrievers ($K=4$)  and $\beta$, a regularizing parameter, is set to 4.

\subsection{Domain Adaptation of Dense L1 Retrievers}
\label{appendix:l1_finetuning}
We used contrastive learning\cite{hardmining} to fine tune the dense retriever to the ObliQA dataset. This involves iteratively fine tuning the encoder with triplet loss\cite{schroff2015facenet}. At the start of each iteration, top-$K$ passages are retrieved and the distractors are selected from the non-GT retrievals which neighbor GT for smooth gradient optimization. Subsequently, training is run for $b$ batches. At the end of the iteration, the fine tuned model is used to retrieve top-$K$ and the process repeats. We used $k=10$, $b=400$ batches and it repeats $n=200$ times, resulting in $8E+4$ training steps. We used a batch size of 8 samples. This follows standard practice as suggested in \cite{hardmining, shrivastavaCVPR16ohem}.

We tried two retrievers: MPNet and E5. Finetuning E5 improved $Recall@10$ from 0.71 to 0.79 \& $MAP@10$ from 0.56 to 0.61. Our eventual system uses E5 as a dense retriever directly and MPNet in the Q2Q retriever module. Using E5 in Q2Q is expected to give better performance but was not evaluated. 

\subsection{Ablation Study of L1 Retrieval Methods}
\label{appendix:ablation}
We performed an exhaustive analysis, trying all possible combinations ($15$) for four different types of retrievers, each bringing into play a different approach for domain adaptation. The results of this analysis are shared in Table \ref{table:l1_ablation}. 

Note that for singleton retrievers, Q2Q is the weakest at $~35\%$, and the domain adapted E5 is the strongest at $~79\%$. The sparse retriever, BM25, lags behind the dense retrievers by $3\%-4\%$ in performance. Adding BM25 or BGE improves the performance of E5 by around $1\%-1.5\%$. Fusing the results of the top three best performing models boosts the performance to $81.89\%$. Adding the Q2Q model achieves a differential gain of $~1.5\%$ giving a final performance of $83.2\%$, leveraging Q-to-Q coherence for retrieving (memorized) information to similar questions encountered by the system in the past.  
    
\begin{table*}[htb]
    \centering
    \begin{tabular}{|l|l|l|l|l|l|l|}
    \hline
        Algorithm & $Recall@10$ & $Recall@20$ & $Recall@40$ & $MAP@20$ & $MAP@10$ & $MAP@40$ \\ \hline \hline
        BM25,Q2Q,BGE,E5-FT & \textbf{0.832} & 0.8822 & 0.9172 & 0.6963 & 0.6914 & 0.6984 \\ \hline
        Q2Q,BGE,E5-FT & 0.8288 & 0.8822 & \textbf{0.9198} & 0.6539 & 0.6487 & 0.6559 \\ \hline
        BM25,Q2Q,E5-FT & 0.8276 & 0.8766 & \textbf{0.9173} & 0.6689 & 0.6641 & 0.6712 \\ \hline
        BM25,Q2Q,BGE & 0.8191 & 0.8616 & 0.9041 & 0.6594 & 0.6552 & 0.6616 \\ \hline
        BM25,BGE,E5-FT & \textbf{0.8189} & 0.8621 & 0.8933 & 0.6935 & 0.6895 & 0.6952 \\ \hline \hline
        BGE,E5-FT & 0.8165 & 0.8655 & 0.895 & 0.6654 & 0.6609 & 0.6671 \\ \hline
        BM25,E5-FT & 0.8108 & 0.8576 & 0.889 & 0.6832 & 0.6789 & 0.6849 \\ \hline
        Q2Q,E5-FT & 0.7968 & 0.8654 & 0.9158 & 0.5263 & 0.5199 & 0.5289 \\ \hline
        BM25,BGE & 0.797 & 0.8341 & 0.8653 & 0.6629 & 0.6593 & 0.6644 \\ \hline
        Q2Q,BGE & 0.7873 & 0.8562 & 0.9089 & 0.5337 & 0.5274 & 0.5364 \\ \hline
        BM25,Q2Q & 0.7798 & 0.8422 & 0.9005 & 0.555 & 0.5493 & 0.5579 \\ \hline \hline
        BGE & 0.7796 & 0.8228 & 0.8564 & 0.6215 & 0.6178 & 0.6233 \\ \hline
        E5-FT & 0.7926 & 0.8446 & 0.8845 & 0.6196 & 0.6148 & 0.6217 \\ \hline
        BM25 & 0.7611 & 0.8022 & 0.8348 & 0.6272 & 0.6236 & 0.6288 \\ \hline
        Q2Q & 0.3539 & 0.5896 & 0.787 & 0.1904 & 0.1729 & 0.1993 \\ \hline
    \end{tabular}
    \caption{Retriever performance for different combinations of L1 retrievers. For more than one retriever, we use RRF to fuse the retrieval results. The table is sorted for $Recall@10$.}
    \label{table:l1_ablation}
\end{table*}

\subsection{Qualitative Analysis of Retrieval Failure Cases}
\label{experiment:false_positive}

An ideal retrieval scenario is where there is a unique hit for each query. In other words, the GT retrieval (containing all relevant information to answer the question) is highly correlated with the query while all other information in the knowledge store isn't. While there are several reasons to consider this an unrealistic assumption, it is important to investigate the chunks which are more correlated to the query than the GT chunk and the reasons thereof.

\begin{figure}[h]
    \centering
    \includesvg[width=0.9\linewidth]{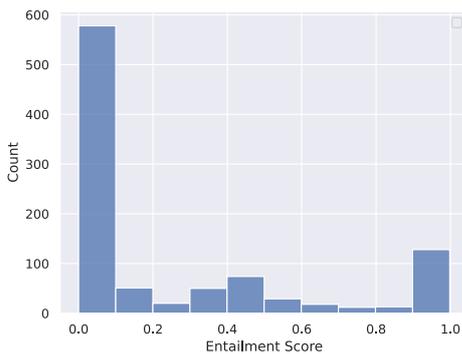}
    \caption{Distribution of entailment scores, $E_s$ (using NLI Deberta v3 Large), for $Top@10$ non-GT retrievals.}
    \label{fig:entailment_hist}
\end{figure}

To do this analysis. we identified all non-GT $Top@10$ retrievals where the contradiction score is low ($C_s < 0.2$). Arguably, these chunks should be correlated with the query, with a low contradiction score, and potentially a high entailment score. In Figure~\ref{fig:entailment_hist}, we plot a histogram of entailment scores for these chunks. A strong mode at $(0<=E_s<0.1)$ is heartening as it shows that NLI can reject non-entailed but highly correlated retrievals. A small peak around $0.5$ may indicate model confusion. However, the peak for the bucket- $(0.9<E_s<=1)$ is intriguing- it seems to indicate the presence of \lq duplicates\rq in the datasets carrying similar information as the GT. The retriever is correctly retrieving these chunks but is getting wrongly penalized. 

We show five examples below which range from having almost duplicate wordings between the GT passage and another $Top@10$ retrieval, to having significant overlap or sharing of phrases/ keywords. 
\begin{itemize}
    \item Table~\ref{table:qualitative_1} shows passages from \textbf{same} document which are near duplicate.
    \item Table~\ref{table:qualitative_2} shows passages from \textbf{different} document which are near duplicate.
    \item Tables~\ref{table:qualitative_31}, \ref{table:qualitative_32}, \ref{table:qualitative_4} show  passages from \textbf{same} document with significant overlap.
\end{itemize}

These examples show that perhaps taking a binary view of retrieval where the retriever is penalized for not retrieving the passage marked as ground truth to be the top-ranked retrieval is perhaps not a proper metric to (a) evaluate, and (b) improve the performance of retrievers.  

\begin{table}[htb!]
    \centering
    \begin{tabular}{|l|l|l|l|l|}
    \hline
        Algo & RePASs & $E_s$ & $C_s$ & $OC_s$ \\ \hline
        RePASs-0 & 0.41 & 0.215 & 0.091 & 0.129 \\ \hline
        RePASs-3 & 0.49 & 0.362 & 0.029 & 0.137 \\ \hline
    \end{tabular}
    \caption{Answer generation performance of the Llama model on the ObliQA dataset. Comparison of metrics with different reasoning context sizes (N).}
    \label{table:n-chunk-repass-score}
\end{table}

\begin{table}[!hb]
    \centering
    \begin{tabular}{|p{0.18\linewidth}|p{0.75\linewidth}|}
    \hline
        Type &  Text \\ \hline
        {\em Ground Truth} (Part 4.40.(7), 17)  & The Regulator may require an Applicant to provide information which the Applicant is required to provide to it under this section in such form, or to verify it in such a way, as the Regulator may direct.\\ \hline
        {\em Non-GT Retrieval}. (Part 11.Chapter 2.108.(3), 17) & The Regulator may require an Applicant to provide information which the Applicant is required to provide to it under this section in such form, or to verify it in such a way, as the Regulator may direct.\\ \hline
    \end{tabular}
    \caption{Near duplicate passages from the same document.}
    \label{table:qualitative_1}
\end{table}

\begin{table}[!hb]
    \centering
    \begin{tabular}{|p{0.18\linewidth}|p{0.72\linewidth}|}
    \hline
        Type & Text \\ \hline
        {\em Ground Truth} (6.6.13. Guidance, 12) & "If a Return is not submitted by the date on which it becomes due \& the Person is in breach of a Rule and the Regulator is entitled to take action including \& but not limited to \& taking steps to withdraw authorisation to conduct Regulated Activities."\\ \hline
        {\em Non-GT Retrieval}. (2.3.8. Guidance, 13) & If a return is not submitted by the date on which it becomes due, the Person is in breach of a Rule and the Regulator is entitled to take action including, but not limited to, taking steps to withdraw authorisation to conduct Regulated Activities.\\ \hline
    \end{tabular}
    \caption{Near duplicate passages from different documents.}
    \label{table:qualitative_2}
\end{table}

\begin{table*}[!hb]
    \centering
    \begin{tabular}{|p{0.18\linewidth}|p{0.78\linewidth}|}
    \hline
        Type &  Text \\ \hline
        {\em Ground Truth} (2.4.2, 9) & "Recognised Bodies. Unless otherwise stated in these Islamic Finance Rules \& a Recognised Body will be entitled to carry on all \& or any part \& of its business as Islamic Financial Business provided that: (a)        it has complied with all other applicable provisions of the Rulebooks issued by the Regulator in relation to the part of its business to be carried on as Islamic Financial Business; and (b)        the carrying on of such part of its business as an Islamic Financial Business has been approved by its Shari'a Supervisory Board." \\ \hline
        {\em Non-GT Retrieval}. (2.4.2. Guidance.(i), 9) & Whether or not all, or any part, of a Recognised Body’s business is to be carried on as Islamic Financial Business, that business must be carried out in compliance with all other relevant parts of the Rulebooks issued by the Regulator.\\ \hline
    \end{tabular}
    \caption{Significant Overlap (Example 1)}
    \label{table:qualitative_31}
\end{table*}

\begin{table*}[!hb]
    \centering
    \begin{tabular}{|p{0.18\linewidth}|p{0.78\linewidth}|}
    \hline
        Type &  Text \\ \hline
        {\em Ground Truth} (D.6., 36) & "Principle 6 – Incorporation of climate-related financial risks into capital and liquidity adequacy processes. Relevant financial firms should incorporate material climate-related financial risks in their internal capital and liquidity adequacy assessment processes." \\ \hline
        {\em Non-GT Retrieval}. (D.6.2., 36) & Principle 6 – Incorporation of climate-related financial risks into capital and liquidity adequacy processes. As part of their internal capital and liquidity adequacy assessment processes, relevant financial firms should consider climate-related financial risks that may impact their capital and liquidity positions over relevant time horizons (e.g., through their impact on traditional risk categories).\\ \hline
    \end{tabular}
    \caption{Significant Overlap (Example 2)}
    \label{table:qualitative_32}
\end{table*}

\begin{table*}[!hb]
    \centering
    \begin{tabular}{|p{0.18\linewidth}|p{0.78\linewidth}|}
    \hline
        Type &  Text \\ \hline
        {\em Ground Truth} (1.2.4. Guidance.1., 4) & "The amount of any supplementary fee will  reflect the Regulator’s reasonable estimate of the  additional time \& effort and resources \& including those of third parties \& necessary to address an issue.  Matters which may cause the Regulator to require the payment of a supplementary fee could include \& for example: a.  complex applications by reason of the Applicant's start-up profile \& origin \& ownership structure or proposed business model; b.  cases where it may be necessary to conduct intense supervisory scrutiny of an entity or individual from a risk perspective; c.  complex restructurings or changes in an Authorised Person’s or Recognised Body's structure or activities; d.  waiver or modification requests which are particularly complex or novel \& in the opinion of the Regulator; e.  novel proposals and applications that cover untested ground or untested areas of the financial services regulatory regime; or f.  assessing complex business models \& the supervision of which will require the Regulator to incur material additional expenses \& such as \& but not limited to \& businesses which operate in \& or rely upon activities performed in jurisdictions with which \& in the view of the Regulator \& insufficient arrangements for co-operation exist between the Regulator and the relevant Non-ADGM Financial Services Regulator(s) in that jurisdiction(s)." \\ \hline
        {\em Non-GT Retrieval}. (1.2.4, 4) & "Supplementary fees The Regulator may require a Person to pay to the Regulator a supplementary fee in circumstances where it expects to incur substantial additional costs or expend substantial additional effort in dealing with an application \& authorisation \& filing or when conducting on-going supervision."\\ \hline
    \end{tabular}
    \caption{Significant Overlap}
    \label{table:qualitative_4}
\end{table*}

\subsection{RePASs Metric and Optimality}
\label{optimal-repass}

The RePASs metric was defined in \citealt{gokhan2024regnlp} as

\begin{eqnarray}
\displaystyle
RePASs = \frac{ \text{E}_{s} - \text{C}_{s} + \text{OC}_{s} + 1}{3} \\
\displaystyle
E_s = \frac{1}{N} \sum_{i=1}^{N} \max_{j} P_{\text{entailment}}(p_j, a_i) \\
\displaystyle
C_s = \frac{1}{N} \sum_{i=1}^{N} \max_{j} P_{\text{contradiction}}(p_j, a_i)
\end{eqnarray}

where \( N \) is the number of sentences in the generated answer, \( P_{\text{entailment(contradiction)}}(p_j, a_i) \) denotes the probability that the \( i \)-th sentence of the answer (\( a_i \)) is entailed (contradicted) by the \( j \)-th sentence of the retrieved passage (\( p_j \)), and \( \max_j \) identifies the maximum probability for each answer sentence among all sentences in the retrieved passage.

\begin{equation}
\displaystyle
OC_s = \frac{1}{M} \sum_{k=1}^{M} \mathbb{1} \left( \max_{l} P_{\text{entailment}}(o_k, a_l) > 0.7 \right)
\end{equation}

where \( M \) is the number of obligation sentences in the retrieved passage, \( P_{\text{entailment}}(o_k, a_l) \) denotes the probability that the \( k \)-th obligation sentence from the passage (\( o_k \)) is entailed by the \( l \)-th sentence in the answer (\( a_l \)), and the indicator function \( \mathbb{1} \) outputs 1 if the maximum entailment score surpasses 0.7, signifying that the obligation is covered.

We now show that under reasonable simplifying assumptions, the RePASs metric can attain the maximal value for a trivial answering model. The following two assumptions imply that the regulatory corpus is reasonable and doesn't contain material that is self-contradictory. 

\begin{assumption}\label{assumption:1}
A sentence from a regulatory corpus entails and doesn't contradict itself, i.e. $P_{\text{entailment}}(p_i,p_i)=1$ and $P_{\text{contradiction}}(p_i,p_i)=0$.
\end{assumption}

\begin{assumption}\label{assumption:2}
Given necessary (maybe unknown) contexts, different sentences across a regulatory corpus should not contradict each other across, so $P_{\text{contradiction}}(p_i,p_j|\text{context}_i,\text{context}_j)=0$ and $P_{\text{contradiction}}(p_j,p_i|\text{context}_j,\text{context}_i)=0$.
\end{assumption}

Using the above assumptions, we can make the following claim (subject to the availability of relevant contexts. Strictly speaking, it changes the definition of the RePASs metric but in a way that we keep to its intended use. 
\begin{claim}
{\bf Passage Concat}, a trivial answering model, which passes through the retrieved passages concatenating them, attains the maximum RePASs score of 1. 
\end{claim} 
\begin{proof}
The Proof follows from the following statements:
\begin{itemize}
    \item ${E}_{s}=1$ since by Assumption~\ref{assumption:1}, for all answer sentences $a_i$, $P_\text{entailment}(a_i,a_i)=1$.
    \item ${C}_{s}=0$ since by Assumption~\ref{assumption:2}, since no two sentences in a regulatory corpus can contradict each other (given the necessary context): $P_\text{contradiction}(a_i,a_j|\text{context}_i,\text{context}_j)=0$.
    \item The {\bf Passage Concat} model covers all context sentences and by Assumption~\ref{assumption:1}, entails them. Hence, $\text{OC}_s=1$.  
\end{itemize}  
\end{proof}

Let's now consider what the above means for practical systems. 

Assumption~\ref{assumption:1} requires that (a) the sentences in a regulatory corpus be meaningful and well formatted; and, (b) an NLI model should have the property that when the premise and hypothesis are exactly the same, it gives an entailment score of 1. While there are reasonable expectations, we see that $\text{E}_s$ is not $1$, though agreeably reasonably high at $0.986$, for the {\bf Passage Concat} model in Table \ref{table:ans_gen_exp}. The fact that the answer from Llama model gives such a poor entailment score ($0.215$) points towards (a) issues due to paraphrasing, perhaps mixing facts from different sentences, and (b) the inability of the NLI model to handle such complex situations. 

Assumption~\ref{assumption:2} is expected to be harder to meet for practical systems. While the non-self-contradiction requirement is reasonable, RsPASs metric requires the property to hold for all sentence pairs. Further, an NLI model may require an unknown context for an NLI model to make such a deduction. As shown in Table \ref{table:ans_gen_exp}, all the answer generation strategies - trivial ones as well as Llama at contradiction scores in the range $0.07-0.1$ (without the appropriate context). We show some examples of contradicting sentences across passages in Appendix \ref{appendix:contradiction}.

\subsection{Self-Contradicting Sentences in ObliQA?}
\label{appendix:contradiction}
In Table \ref{table:contradiction-example-1}, \ref{table:contradiction-example-2}, \ref{table:contradiction-example-3}, we show three example pairs containing highlighted sentence pairs having contradiction scores (using NLI Deberta v3 Large) of at least $0.5$. While these examples may not be particularly illuminating, they do point to the challenges in NLI and the potential complexities in the real world data. In all these cases, the NLI model should've slotted them into the $NEI$ class.

Passage 1 (PART 4.9.1.1.Application.Guidance.1 of Document 6) and Passage 2 (PART 3.7.1.1 of Doc 6 ) in Table \ref{table:contradiction-example-1} refer to the obligatory requirements for a 'Foreign Fund Manager'. It is unreasonable to deduce that the highlighted sentences present any kind of contradiction. 

\begin{table*}[!hb]
    \centering
    \begin{tabular}{|p{0.18\linewidth}|p{0.70\linewidth}|} \hline Type & Text  \\ \hline
        Passage 1 &refer to 
        \hl{A Foreign Fund Manager must also comply with the requirements in this Chapter, because it is managing a Domestic Fund.} \\ \hline
        Passage 2 & \hl{Subjecting to the Abu Dhabi Global Market jurisdiction.} A Foreign Fund Manager to whom this Chapter applies must: (a) be subject to regulation by, or registration with, a Financial Services Regulator in a ..... \\ \hline
    \end{tabular}
    \caption{Contradiction Example - 1}
    \label{table:contradiction-example-1}
\end{table*}

Similarly in the scenario in Table~\ref{table:contradiction-example-2} where the NLI model shows high contradiction between passage 97 of Doc 31 and passage 4) of Doc 31, and 

Table~\ref{table:contradiction-example-3} showing between passage APP 4.50. of Doc 11 and 9.2.7 of Doc 11. These short phrases/ titles can't be deemed to indicate contradictions, and point to shortcomings of the NLI model.

\begin{table*}[!hb]
    \centering
    \begin{tabular}{|p{0.18\linewidth}|p{0.70\linewidth}|}
    \hline
        Type & Text  \\ \hline
        Passage 1 & \hl{INTERACTION OF CHAPTER 12 WITH OTHER RULE DISCLOSURE OBLIGATIONS.} Offers, and Admission to FSRA Official List of Securities Considering the circumstances above, and the positioning of the FSRA in relation to these matters, the FSRA suggests that Issuers/Petroleum Reporting Entities (and their advisors) contact the FSRA as early as possible to discuss. \\ \hline
        Passage 2 & \hl{INTRODUCTION.} In the context of the obligations and disclosures by Petroleum Reporting Entities, the FSRA operates as the Listing Authority within ADGM and is therefore charged with supervising Petroleum Reporting Entity disclosures under FSMR, MKT and by incorporation in Chapter 12 of MKT, the PRMS. \\ \hline
    \end{tabular}
    \caption{Contradiction Example - 2}
    \label{table:contradiction-example-2}
\end{table*}

\begin{table*}[!hb]
    \centering
    \begin{tabular}{|p{0.18\linewidth}|p{0.70\linewidth}|}
    \hline
        Type & Text  \\ \hline
        Passage 1 & \hl{Audit committee.} A separate section of the annual report should describe the work of the audit committee in discharging its responsibilities. The annual report should also
explain to Shareholders how, if the auditor provides non audit services, auditor objectivity and independence is safeguarded. Principle 5 - Shareholder rights and effective dialogue Rule 9.2.7 ....... \\ \hline
        Passage 2 & \hl{Principle 5 - Shareholder rights and effective dialogue}. The Board must ensure that the rights of Shareholders are properly safeguarded through appropriate measures that enable the Shareholders to exercise their rights effectively, promote effective dialogue with Shareholders and other key stakeholders as appropriate, and prevent any abuse or oppression of minority Shareholders. \\ \hline
    \end{tabular}
    \caption{Contradiction Example - 3}
    \label{table:contradiction-example-3}
\end{table*}

\subsection{Prompts}
\label{appendix:prompts}

\noindent \textbf{RegNLP Answer Generation Prompt}
\noindent Quoted from \citealt{gokhan2024regnlp}
\begin{lstlisting}
You are a regulatory compliance assistant. Provide a detailed answer for the question that fully integrates all the obligations and best practices from the given passages. Ensure your response is cohesive and directly addresses the question. Synthesize the information from all passages into a single, unified answer.

question: {question} 
passages: {context} 
answer:
\end{lstlisting}

\noindent \textbf{LLM as a Judge Prompt}
\noindent Quoted from \citealt{roucherllmjudge}

\begin{lstlisting}
You will be given a user_question and system_answer couple.
Your task is to provide a 'total rating' scoring how well the system_answer answers the user concerns expressed in the user_question.
Give your answer on a scale of 1 to 4, where 1 means that the system_answer is not helpful at all, and 4 means that the system_answer completely and helpfully addresses the user_question.

Here is the scale you should use to build your answer:
1: The system_answer is terrible: completely irrelevant to the question asked, or very partial
2: The system_answer is mostly not helpful: misses some key aspects of the question
3: The system_answer is mostly helpful: provides support, but still could be improved
4: The system_answer is excellent: relevant, direct, detailed, and addresses all the concerns raised in the question

Provide your feedback as follows:

Feedback:::
Evaluation: (your rationale for the rating, as a text)
Total rating: (your rating, as a number between 1 and 4)

You MUST provide values for 'Evaluation:' and 'Total rating:' in your answer.

Now here are the question and answer.

Question: {question}
Answer: {answer}

Provide your feedback.
Feedback:::
Evaluation: 
\end{lstlisting}

\end{document}